\documentclass[twocolumn]{aastex6}
\usepackage{autoaffil}
\usepackage{mathtools}
\usepackage{CJKutf8}
\usepackage{hyperref}

\makeatletter
\renewcommand\plotone[1]{% 
 \centering 
 \leavevmode 
 \setlength{\plot@width}{\linewidth}
 \includegraphics[width={\eps@scaling\plot@width}]{#1}% 
}
\renewcommand\plottwo[2]{% 
 \centering 
 \leavevmode 
 \setlength{\plot@width}{0.5\linewidth}
 \includegraphics[width={\eps@scaling\plot@width}]{#1}% 
 \hfil 
 \includegraphics[width={\eps@scaling\plot@width}]{#2}% 
}
\makeatother
\hypersetup{citecolor=blue,linkcolor=blue,urlcolor=blue}
\usepackage[hang]{footmisc}

\renewcommand{\textsc}{}
\newcommand{\plotdir}{}
\newcommand{\ra}[4]{$#1\overset{\mathrm h}{\phd}#2\overset{\mathrm m}{\phd}#3\overset{\mathrm s}{.}#4$}
\newcommand{\dec}[4]{$#1\overset{\circ}{\phd}#2\overset{\prime}{\phd}#3\overset{\prime\prime}{.}#4$}
\newcommand{\cn}[1]{~(\begin{CJK*}{UTF8}{gbsn}#1\end{CJK*}\!\!)}
\newcommand{\jp}[1]{~(\begin{CJK*}{UTF8}{min}#1\end{CJK*}\!\!)}

\shorttitle{{\sc Circumstellar Interaction in the Low-Luminosity Type~IIP SN~2016}bkv}
\shortauthors{Hosseinzadeh et al.}

\begin{document}

\title{Short-Lived Circumstellar Interaction in the Low-Luminosity Type~IIP SN~2016\lowercase{bkv}}
\author{
Griffin~Hosseinzadeh\affiliation{\afLCO}\affilcite{LCO,UCSB,DSFP},
Stefano~Valenti\affilcite{UCD},
Curtis~McCully\affilcite{LCO,UCSB},
D.~Andrew~Howell\affilcite{LCO,UCSB},
Iair~Arcavi\affilcite{LCO,UCSB,Einstein},
Anders~Jerkstrand\affilcite{MPA},
David~Guevel\affilcite{LCO,UCSB,HCO},
Leonardo~Tartaglia\affilcite{UCD,UA,OKC},
Liming~Rui\cn{芮黎明}\affilcite{Tsinghua},
Jun~Mo\cn{莫军}\affilcite{Tsinghua},
Xiaofeng~Wang\cn{王晓锋}\affilcite{Tsinghua},
Fang~Huang\cn{黄芳}\affilcite{Tsinghua},
Hao~Song\cn{宋浩}\affilcite{Tsinghua},
Tianmeng~Zhang\cn{张天萌}\affilcite{NAOC},
and Koichi~Itagaki\jp{板垣公一}\affilcite{Itagaki}
}
\affilblock

\received{2017 December 28}
\revised{2018 May 5}
\accepted{2018 May 14}

\begin{abstract}
While interaction with circumstellar material is known to play an important role in Type~IIn supernovae (SNe), analyses of the more common SNe~IIP and IIL have not traditionally included interaction as a significant power source. However, recent campaigns to observe SNe within days of explosion have revealed narrow emission lines of high-ionization species in the earliest spectra of luminous SNe~II of all subclasses. These ``flash ionization'' features indicate the presence of a confined shell of material around the progenitor star. Here we present the first low-luminosity (LL) SN to show flash ionization features, SN~2016bkv. This SN peaked at $M_V = -16$~mag and has H$\alpha$ expansion velocities under 1350~km~s$^{-1}$ around maximum light, placing it at the faint/slow end of the distribution of SNe~IIP (similar to SN~2005cs). The light-curve shape of SN~2016bkv is also extreme among SNe~IIP. A very strong initial peak could indicate additional luminosity from circumstellar interaction. A very small fall from the plateau to the nickel tail indicates unusually large production of radioactive nickel compared to other LL SNe~IIP. A comparison between nebular spectra of SN~2016bkv and models raises the possibility that SN~2016bkv is an electron-capture supernova.
\end{abstract}


\section{Introduction}
\affootnotes

The majority of massive ($M_\mathrm{ZAMS} \gtrsim 8\,M_\sun$) stars end their lives as Type~II supernovae (SNe), i.e., those with hydrogen-rich spectra \citep{Smartt2009,Arcavi2016}. Broadly speaking, SNe~II can be divided into two subclasses. SNe~IIP/L have light curves that plateau (P) or decline linearly (L) for roughly 100 days before falling to a radioactive-decay-powered tail \citep{Barbon1979}. SNe~IIn have spectra dominated by narrow (n) hydrogen emission lines, indicating strong interaction with circumstellar material \citep[CSM;][]{Schlegel1990,Chugai1991}. Hereinafter, we will use ``SNe~II'' to refer to SNe~IIP and IIL as a single class, excluding SNe~IIb and IIn.

SNe~II peak in the range of $-14 \gtrsim M_B \gtrsim -18$~mag \citep{Patat1994} and show a correlation between plateau luminosity and photospheric expansion velocities \citep{Hamuy2002,Pejcha2015}. \cite{Pumo2017} claim that the ratio of the explosion energy to the ejected mass controls the plateau luminosity of SNe~II, where smaller ratios are less luminous (see their Figure~5). At the faint/slow end of this distribution is a group of events referred to as low-luminosity (LL) SNe~II \citep{Pastorello2004,Spiro2014}, exemplified by the nearby, well-studied SN~2005cs, whose red supergiant (RSG) progenitor star was detected by \cite{Maund2005} and \cite{Li2006} in pre-explosion images taken by the \textit{Hubble Space Telescope} (\textit{HST}). Several models have been proposed to explain LL SNe~II, including high-mass RSGs exploding as fallback SNe \citep{Turatto1998,Zampieri1998} and low-mass RSGs exploding as electron-capture SNe \citep{Chugai2000}.

Interaction with CSM traditionally has not been considered significant in SNe~II, since broad P~Cygni lines dominate their spectra. However, recent efforts to obtain spectra of SNe within days of explosion have resulted in the discovery of narrow emission features in a significant fraction of early spectra of SNe~II \citep[18\% of those observed within 5~days of explosion;][]{Khazov2016}. Early examples of these features include SNe~1983K \citep{Niemela1985}, 1993J \citep{Benetti1994,Garnavich1994,Matheson2000a}, 1998S \citep{Leonard2000,Shivvers2015}, and 2006bp \citep{Quimby2007}. \cite{Gal-Yam2014} first coined the term ``flash spectroscopy'' to describe this emission, produced as the CSM recombines after being ionized by the initial shock breakout flash. More recently, \cite{Yaron2017} observed flash spectra of the Type~II SN~2013fs changing on hour time scales.

Intriguingly, \cite{Khazov2016} suggest that the presence of flash ionization features may be correlated with luminosity; in fact, all of their flash ionization spectra are from events with peak absolute magnitudes $M_R < -17.5$~mag. This implies that more luminous events maintain higher temperatures at early times, enabling high-ionization lines to persist in their spectra long enough to be detected.

The object presented here, SN~2016bkv, is an SN~II that was discovered soon after explosion. With a peak absolute magnitude of $M_V = -16$~mag and H$\alpha$ expansion velocities under 1350~km~s$^{-1}$, it greatly resembles SN~2005cs and other LL events in the literature. However, unlike any previous LL SN~II, its pre-maximum spectra are dominated by narrow emission lines from high-ionization species, making it the least luminous SN for which flash ionization of the CSM has been detected. In this article, we report observations of SN~2016bkv and discuss their implications for the role of CSM interaction in SNe~II. We also investigate the progenitor of SN~2016bkv, in particular, whether it exploded due to iron core collapse or electron capture.

\section{Discovery}
\cite{Itagaki2016} discovered SN~2016bkv on 2016 March 21.70 UT at 17.5~mag using an unfiltered Celestron 14~inch (35~cm) telescope. \cite{Ross2016} do not detect the supernova on March 19.29, but only to an unfiltered limiting magnitude of 17~mag. Since this limit is not as deep as the first detection, in order to determine the explosion time, we fit $F(t) = F_1(t-t_0)^2$ to the first three unfiltered photometry points, after which the light curve appears to rise more slowly. This yields March 20.5 ($\mathrm{MJD}_0 = 57467.5 \pm 1.2$) as the explosion epoch, where we have adopted the time between estimated explosion and first detection as the uncertainty. Coincidentally, this uncertainty range includes the entire period from nondetection to first detection. \cite{2016TNSCR.239....1H} obtained the first spectrum on March 23.5, classifying the transient as a young SN~II.

At right ascension \ra{10}{18}{19}{31} and declination \dec{+41}{25}{39}{3} (J2000), SN~2016bkv lies 30$\overset{\prime\prime}{.}$3 (2.11~kpc projected) from the center of the spiral galaxy NGC~3184, which has a redshift of $z = 0.001975 \pm 0.000003$ \citep{Strauss1992} and a Cepheid-based distance modulus of $\mu = 30.79 \pm 0.05$~mag \citep{Ferrarese1999},\footnote{Note, however, that other methods give distance moduli ranging from 29.3~mag \citep{Pierce1994} to 31.2~mag \citep{Jones2009}.} corresponding to a luminosity distance of $d_L = 14.4 \pm 0.3$~Mpc. In the past century, NGC~3184 has hosted at least four other SNe---1921B, 1921C, 1937F \citep{Shapley1939},\footnote{Spectra of these SNe from E.~Hubble's and F.~Zwicky's programs were never obtained, but they have been photometrically classified as Types~II, I, and IIP, respectively \citep{Barbon2008}.} and 1999gi \citep[Type~II;][]{Schlegel2001,Smartt2001}---and one SN impostor---2010dn, which \cite{Smith2011} identify as the outburst of a luminous blue variable star.

As noted by \cite{Milisavljevic2016}, images of NGC~3184 from \emph{HST} show a star cluster at the position of SN~2016bkv. When the SN fades, it may be possible to determine the luminosity and colors of the progenitor star based on the change in flux from the cluster.

\section{Observations and Data Reduction}
In addition to unfiltered photometry from the 35 and 60~cm telescopes at the Itagaki Astronomical Observatory (Yamagata, Japan), we obtained \textit{UBVgri} photometry of SN~2016bkv using Las Cumbres Observatory's \citep{Brown2013} robotic 1~m telescope located at the McDonald Observatory (TX, USA) and \textit{UBVRI} photometry using the 0.8~m Tsinghua University--National Astronomical Observatory of China Telescope \citep[TNT;][]{Huang2012} located at the Xinglong Observatory (Hebei Province, China). Unfortunately, the \textit{Swift} satellite could not observe the SN in ultraviolet due to a nearby bright star.

\begin{figure*}
\plotone{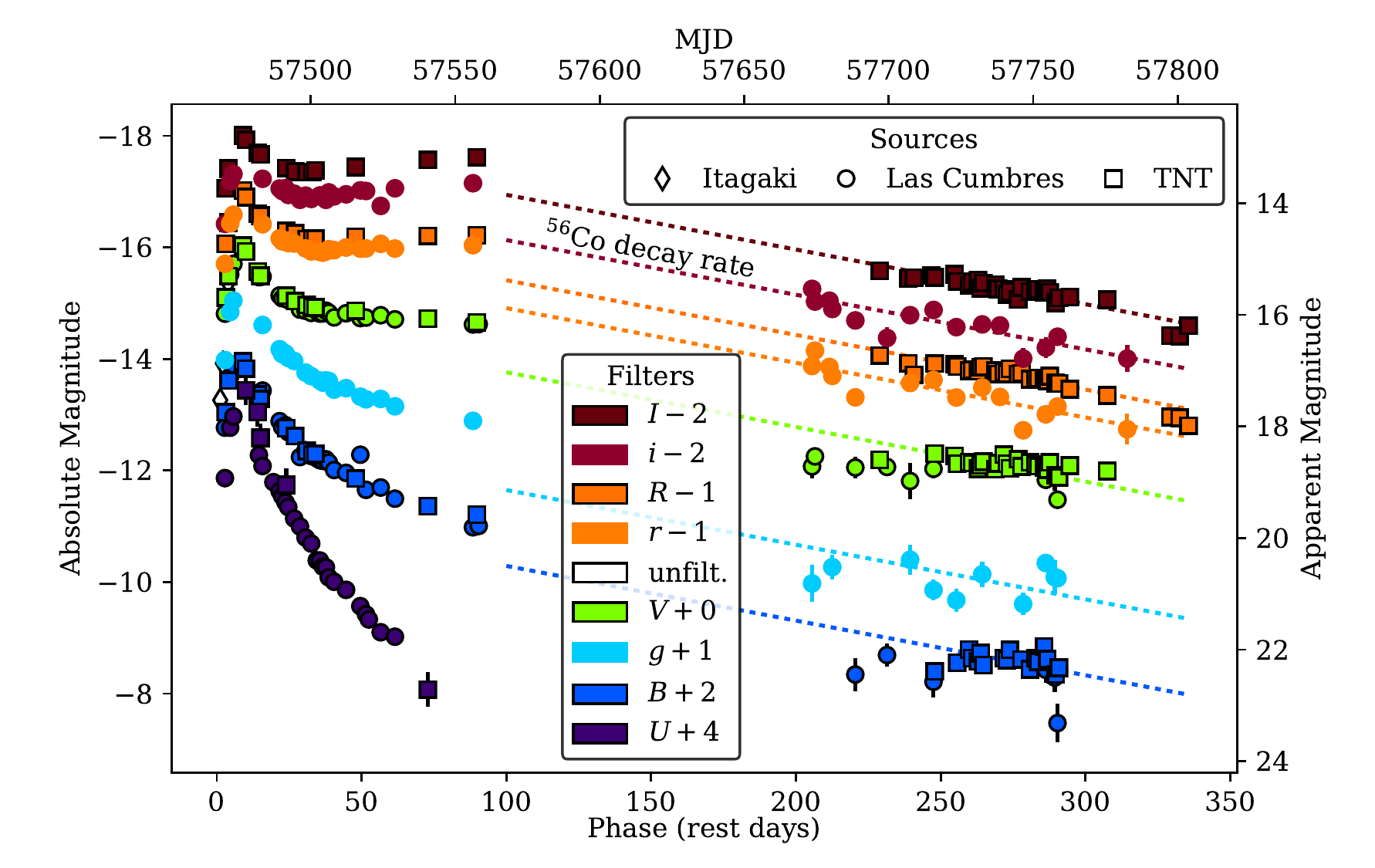}
\caption{\scriptsize One year of photometry of SN~2016bkv. Late nondetections are hidden for clarity. Note the strong early bump and the small fall from plateau, which happened while the supernova was behind the sun. (The data used to create this figure are available.)\label{fig:phot}}
\end{figure*}

Photometry measurements were complicated by the fact that the SN occurred in a bright but unresolved \ion{H}{2} region \citep[number 62 in the list of][]{Hodge1983} and about $7\overset{\prime\prime}{.}7$ away from an even brighter star \citep[SDSS~J101819.82+412544.3;][]{Adelman-McCarthy2007}. We obtained \textit{BVgri} reference images with Las Cumbres approximately 600~days after explosion, when the SN was much fainter than the \ion{H}{2} region, and subtracted them from the images of the supernova using PyZOGY \citep{PyZOGY}, a new open-source Python image subtraction pipeline based on the optimal algorithm of \cite{Zackay2016}. We then measured aperture photometry on the difference images using \texttt{lcogtsnpipe} \citep{Valenti2016}. The $U$-band light curve is PSF photometry on unsubtracted images, because host contamination in our reference images is much weaker in $U$. \textit{UBV} photometry is calibrated in the Vega system to \cite{Landolt1992} standard fields observed on the same night as the SN. \textit{gri} photometry is calibrated in the AB system to stars in the Sloan Digital Sky Survey Data Release 8 \citep[SDSS DR8;][]{Aihara2011}.

We obtained a $U$-band reference image with TNT approximately 300~days after explosion and \textit{BVRI} references approximately 600~days after explosion. We subtracted these from the TNT images of the supernova using HOTPANTS \citep{HOTPANTS} and measured PSF photometry using a custom pipeline based on PyRAF tasks \citep{PyRAF}. TNT \textit{UBVRI} photometry is calibrated in the Vega system to SDSS DR13 stars \citep{Albareti2017}.

We measured PSF photometry from the remaining unsubtracted images using standard PyRAF tasks. Unfiltered photometry is calibrated to 20 nearby stars in SDSS DR6 \citep{Adelman-McCarthy2007} by fitting blackbody spectra to their \textit{ugriz} photometry and, for stars with reasonable fits, convolving the spectra with the spectral response curve of K.~Itagaki's KAF-1001E CCD (both telescopes use the same CCD) to produce AB magnitudes. The resulting unfiltered magnitudes are not relevant to the bolometric light curve we produce in Section~\ref{sec:bolometric}, but we do consider them when fitting the early light curve in Section~\ref{sec:sapir-waxman}. The photometry is plotted in Figure~\ref{fig:phot}.

\begin{figure}
\plotone{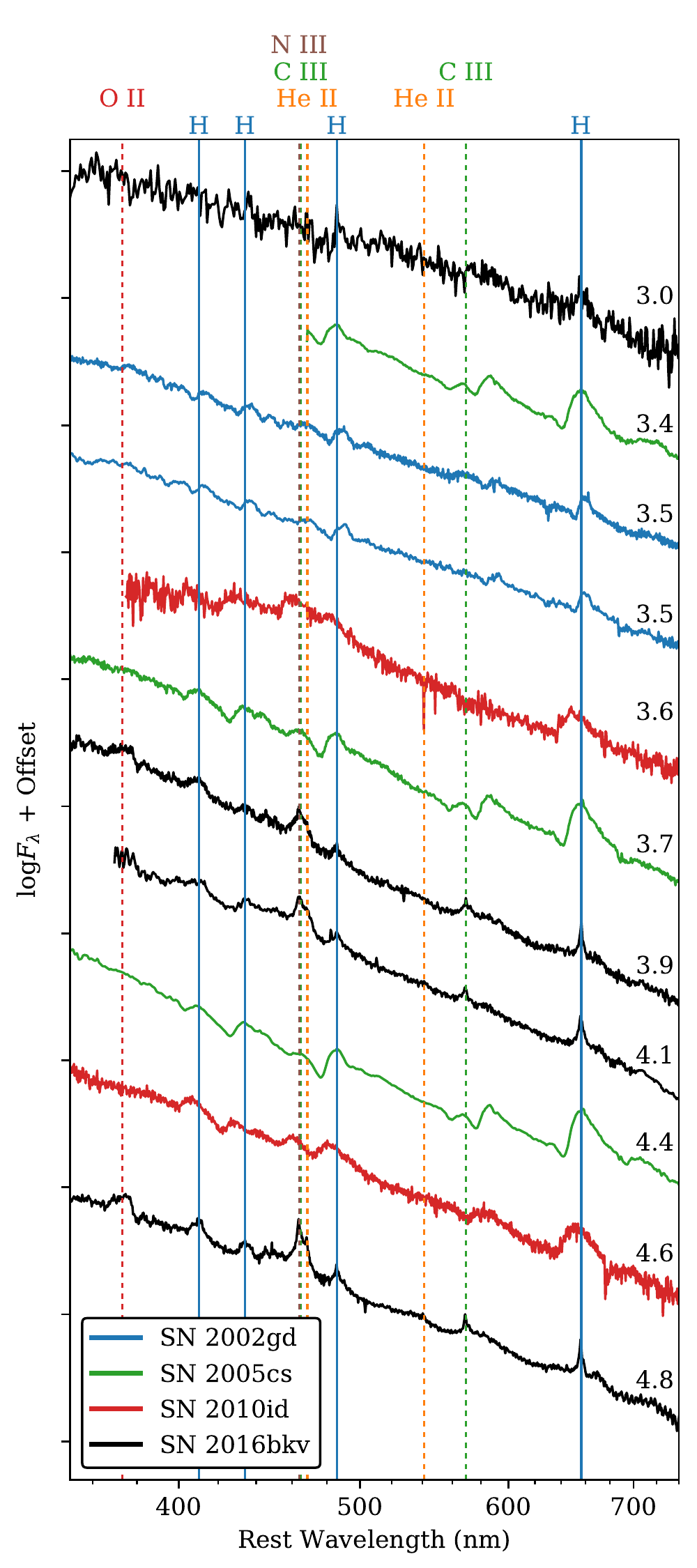}
\caption{\scriptsize Flash-ionized spectra of SN~2016bkv compared with all spectra of LL SNe~II observed within five days of estimated explosion: SNe 2002gd \citep{Faran2014,Spiro2014}, 2005cs \citep{Pastorello2006,Faran2014}, and 2010id \citep{Gal-Yam2011}. Phases in days from estimated explosion are marked to the right of each spectrum. SN~2016bkv is the only one that shows narrow emission of high-ionization species (dotted vertical lines).\label{fig:flashspec}}
\end{figure}

\begin{figure}
\plotone{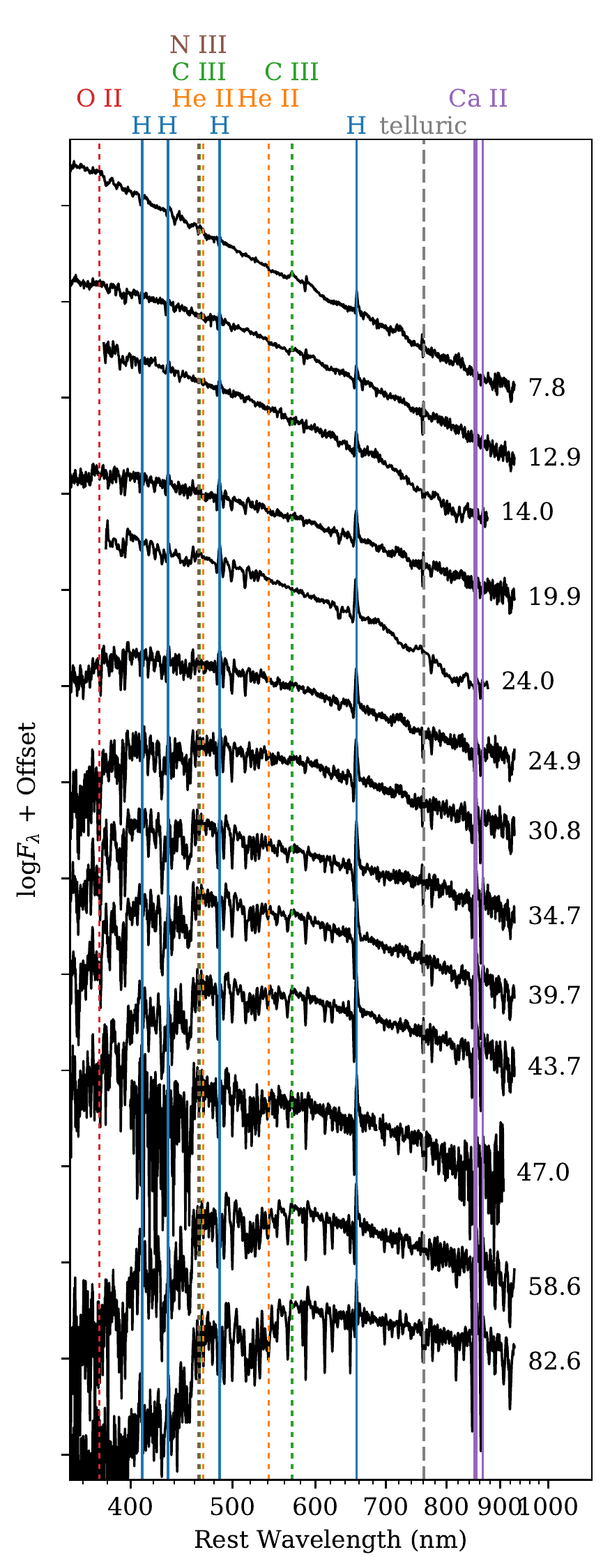}
\caption{\scriptsize Photospheric spectra of SN~2016bkv; see Figure~\ref{fig:synapps} for a more detailed line identification during the photospheric phase. Phases in days from estimated explosion are marked to the right of each spectrum. The early high-ionization lines in Figure~\ref{fig:flashspec} disappear after 5 days, giving way to typical SN~II spectra with very low photospheric velocities (H$\alpha$ velocity $< 1350\mathrm{\,km\,s^{-1}}$).\label{fig:spec}}
\end{figure}

We obtained 17 optical spectra of SN~2016bkv during the flash ionization (Figure~\ref{fig:flashspec}) and photospheric (Figure~\ref{fig:spec}) phases using the robotic FLOYDS instrument, mounted on Las Cumbres Observatory's 2~m telescope on Haleakal\={a} (HI, USA), and the Beijing Faint Object Spectrograph and Camera (BFOSC) and the Optomechanics Research Inc.\ (OMR) low-resolution spectrograph, both mounted on the 2.16~m telescope at Xinglong \citep{Fan2016}. Note that these low-resolution spectrographs cannot resolve lines with velocities below $\sim\!800$~km~s$^{-1}$, meaning that we cannot constrain the photospheric velocity of SN~2016bkv during most of its evolution. Four additional spectra were obtained during the nebular phase using the Kast Double Spectrograph \citep{Miller1994} on the C.~Donald Shane Telescope at the Lick Observatory (CA, USA) and the Low-Resolution Imaging Spectrometer \citep[LRIS;][]{Oke1995,Rockosi2010} on the Keck~I telescope at the W.~M.~Keck Observatory (HI, USA). All spectra are logged in Table~\ref{tab:spec} and available for download from the \href{https://wiserep.weizmann.ac.il}{Weizmann Interactive Supernova Data Repository} \citep{Yaron2012}.

\begin{deluxetable}{CllR}
\tablecaption{Log of Spectroscopic Observations\label{tab:spec}}
\tablehead{\colhead{MJD} & \colhead{Telescope} & \colhead{Instrument} & \colhead{Phase (days)}}
\decimals
\startdata
57470.5 & Las Cumbres 2 m & FLOYDS & 3.0 \\
57471.4 & Las Cumbres 2 m & FLOYDS & 3.9 \\
57471.6 & Xinglong 2.16 m & BFOSC & 4.1 \\
57472.3 & Las Cumbres 2 m & FLOYDS & 4.8 \\
57475.3 & Las Cumbres 2 m & FLOYDS & 7.8 \\
57480.4 & Las Cumbres 2 m & FLOYDS & 12.9 \\
57481.6 & Xinglong 2.16 m & BFOSC & 14.0 \\
57487.4 & Las Cumbres 2 m & FLOYDS & 19.9 \\
57491.6 & Xinglong 2.16 m & BFOSC & 24.0 \\
57492.4 & Las Cumbres 2 m & FLOYDS & 24.9 \\
57498.4 & Las Cumbres 2 m & FLOYDS & 30.8 \\
57502.3 & Las Cumbres 2 m & FLOYDS & 34.7 \\
57507.3 & Las Cumbres 2 m & FLOYDS & 39.7 \\
57511.3 & Las Cumbres 2 m & FLOYDS & 43.7 \\
57514.6 & Xinglong 2.16 m & OMR & 47.0 \\
57526.3 & Las Cumbres 2 m & FLOYDS & 58.6 \\
57550.3 & Las Cumbres 2 m & FLOYDS & 82.6 \\
57724.6 & Shane   & Kast   & 256.6 \\
57780.4 & Shane   & Kast   & 312.2 \\
57903.3 & Keck I  & LRIS   & 434.9 \\
58075.6 & Keck I  & LRIS   & 606.9 \\
\enddata
\end{deluxetable}

\section{Analysis}
\subsection{Flash Spectroscopy}\label{sec:flashspec}
Initially, the spectrum of SN~2016bkv shows narrow emission lines of hydrogen (410.1, 434.0, 486.1, and 656.3~nm), \ion{He}{2} (468.6 and 541.1~nm), \ion{C}{3} (464.8 and 569.6~nm), \ion{N}{3} (464.0~nm), and possibly \ion{O}{2} (371.3--374.9~nm). With the exception of \ion{O}{2}, these lines are a subset of those identified by \cite{Gal-Yam2014} in the Type~IIb SN~2013cu \citep[modeled by][]{Groh2014} and by \cite{Yaron2017} in the Type~II SN~2013fs. The line widths are likely unresolved by our low-resolution spectrographs.

Previously, only three LL SNe~II were observed spectrally this soon after explosion: SNe 2002gd \citep{Faran2014,Spiro2014}, 2005cs \citep{Pastorello2006,Faran2014}, and 2010id \citep{Gal-Yam2011}. Although their spectra are similarly blue, none of them show narrow emission features, instead being dominated by the same P~Cygni profiles as during the photospheric phase (see Figure~\ref{fig:flashspec}). Notably, \cite{Gal-Yam2014} identify a feature around 468~nm as \ion{He}{2}, an indicator of very high temperatures, in their second spectrum of SN~2010id. However, \cite{Pastorello2004} identify a similar feature as high-velocity H$\beta$ in their spectra of SN~2005cs (see their Figure~5).

Because of the very low photospheric velocity of SN~2016bkv, lower than these other three events, we suggest that its CSM may have lasted longer before being swept up by supernova ejecta. This would have made flash spectroscopy easier by allowing us more time after explosion to observe the narrow emission lines. A larger sample size of well-observed flash-ionized SNe~II would allow us to search for a correlation between photospheric velocity and duration of flash spectra.

\subsection{Blackbody Fitting and Bolometric Light Curve}\label{sec:bolometric}
\begin{figure}
\plotone{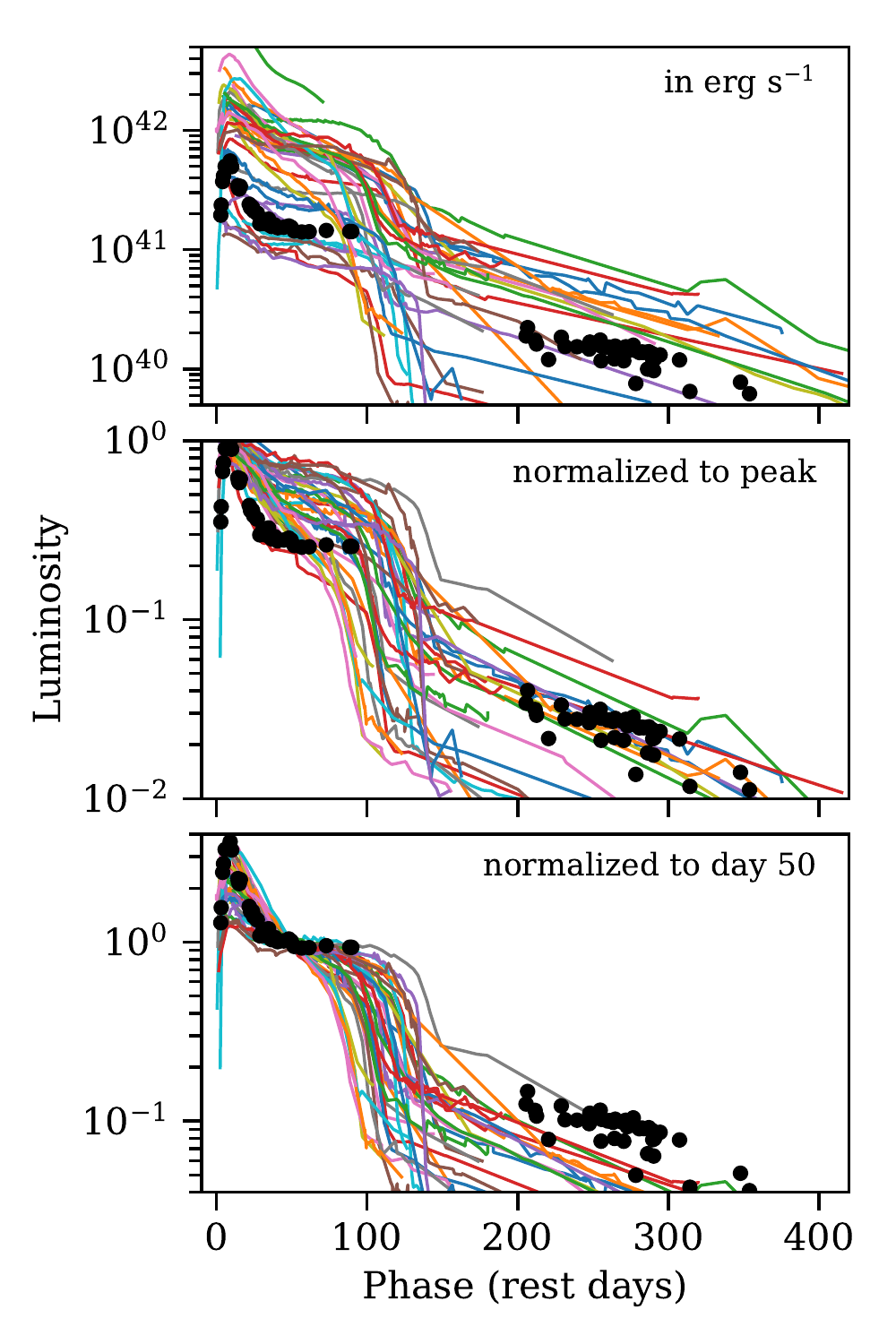}
\caption{Pseudobolometric light curve of SN~2016bkv (black dots) compared to the \cite{Valenti2016} sample of SNe~II (colored lines). SN~2016bkv has among the most extreme initial decline slopes and the smallest plateau-to-tail luminosity ratios.\label{fig:bolo}}
\end{figure}

We construct a bolometric light curve of SN~2016bkv by fitting each epoch of photometry with a blackbody spectrum. In order to compare to pseudobolometric light curves from the literature, typically constructed using only optical photometry, we then integrate the blackbody spectrum from $U$ to $I$. The result is shown in Figure~\ref{fig:bolo}, where it is compared to pseudobolometric light curves of other SNe~II from \cite{Valenti2016}.

The pseudobolometric light curve of SN~2016bkv is extreme in two ways. First, it has a very strong peak around 7~days after explosion with respect to the day-50 plateau luminosity ($L_{50}$). Integrating the light curve over the first 40 days, we find an excess $\Delta E = \int_0^{40} (L - L_{50}) dt = 4.3 \times 10^{47}\,\mathrm{erg}$ above the plateau. If all of this energy were the result of converting ejecta kinetic energy into light with 100\% efficiency, the mass of ejecta stopped would have to be
\[ M = \frac{2 \Delta E}{v_\mathrm{ej}^2} = 0.04\,M_\sun \left(\frac{1000\,\mathrm{km\,s}^{-1}}{v_\mathrm{ej}}\right)^2, \]
and the CSM mass would have to be comparable \citep{Smith2016}. For comparison, \cite{Yaron2017} estimate that an order of magnitude less, a few $10^{-3}\,M_\sun$, CSM was necessary to produce the flash ionization lines in the spectra of SN~2013fs. However, the light-curve peak of SN~2016bkv may be powered by a combination of shock cooling and circumstellar interaction, in which case less CSM would be required.

Second, the luminosity ratio between the plateau (day 50) and the radioactive nickel tail (day 200) is relatively small. This implies that the amount of nickel produced in the explosion is unusually large for an LL SN~II. Comparing the luminosity on the tail to SN~1987A, as in \cite{Hamuy2003}, we find that SN~2016bkv produced $M_\mathrm{Ni} = 0.0216 \pm 0.0014 \,M_\sun$ of nickel. The uncertainty in this measurement is almost entirely due to the 2\% uncertainty on the distance to the supernova, the 5\% assumed uncertainty on the host-galaxy extinction ($A_V = 0.00 \pm 0.05$~mag), and the 1.2\% uncertainty from the explosion epoch. 

This measurement is surprisingly higher than those for previous LL SNe~IIP and merits a careful consideration of the uncertainties. If supernova light was still present in the reference images we used to create our subtracted light curve, the luminosity we derived would be lower than the true luminosity, making the nickel mass even higher. Likewise, if extinction from dust in the ejecta were important, the true underlying luminosity would be higher than we observe, again making the nickel mass higher. In order to obtain a nickel mass similar to those of other LL SNe~IIP, the true luminosity would have to be about three times lower than what we observe.

\begin{figure}
\plotone{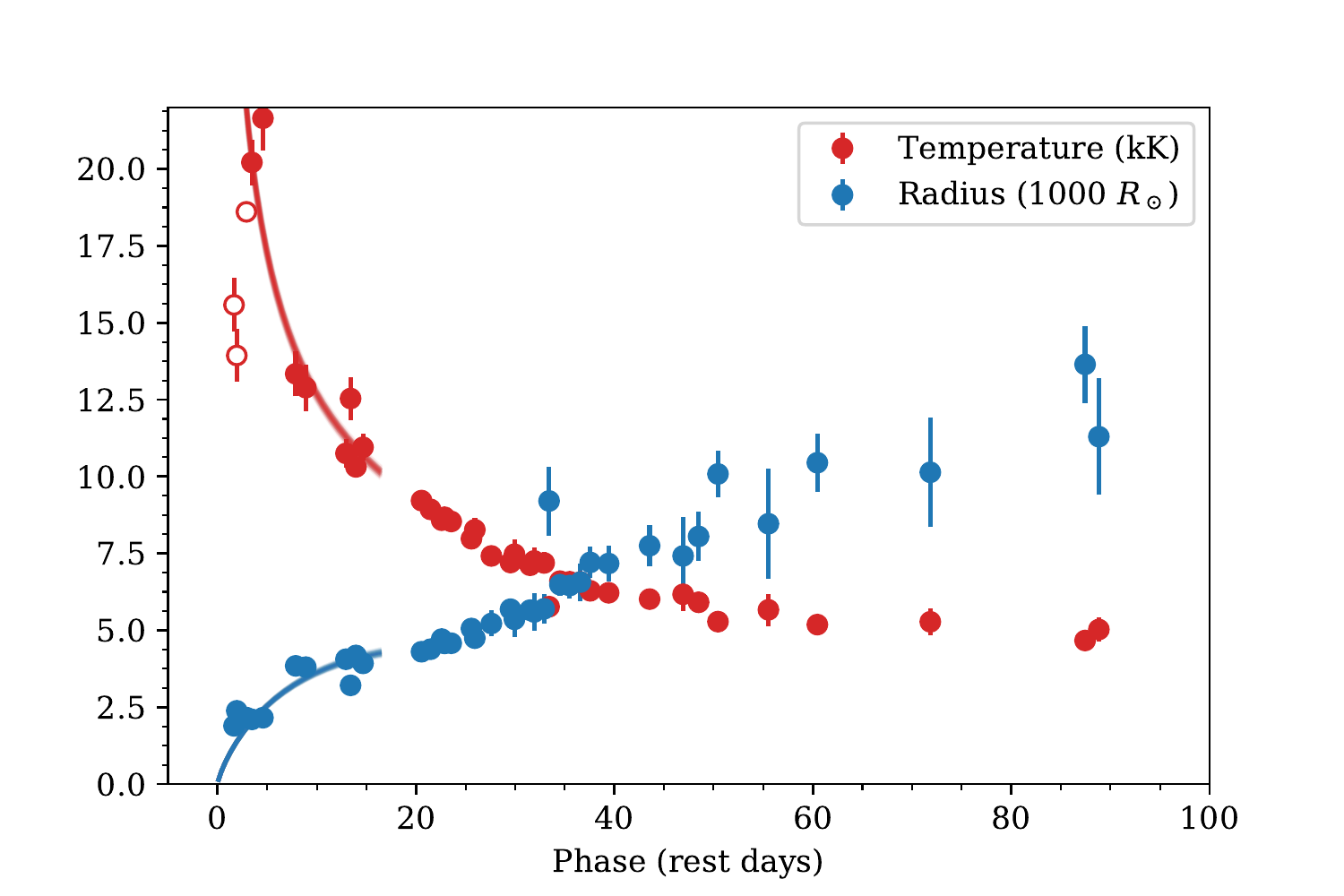}
\caption{Best-fit blackbody temperature (red points) and radius (blue points) of SN~2016bkv during the photospheric phase, compared to the temperature (red lines) and radius (blue lines) from the best-fit \cite{Sapir2017} models in Section~\ref{sec:sapir-waxman}. The open red points are likely underestimates of the temperature from when the SED peaks blueward of our observations.\label{fig:bb}}
\end{figure}

Figure~\ref{fig:bb} shows the evolution of the blackbody temperature and radius from the fit. The temperature evolution shows an unexpected rise during the first six days. Because we lack photometry blueward of $U$, the blackbody fits are not strongly constrained at these early times, when the spectral energy distribution peaks in the ultraviolet. If this rising behavior is real, it might be related to the flash ionization of the CSM, which manifests itself in the spectra during the same time period. However, given the uncertainties in the fits, we cannot claim to have observed a significant rise in temperature.

\subsection{Shock Cooling Model Fitting}\label{sec:sapir-waxman}
\cite{Sapir2017} present a method for modeling early supernova light curves powered by shock cooling emission---the radiation of energy deposited in the stellar envelope by the core-collapse shock wave. Here we test whether their models alone can reproduce the unusually sharp peak of SN~2016bkv or whether another power source is needed.

\begin{figure*}
\plotone{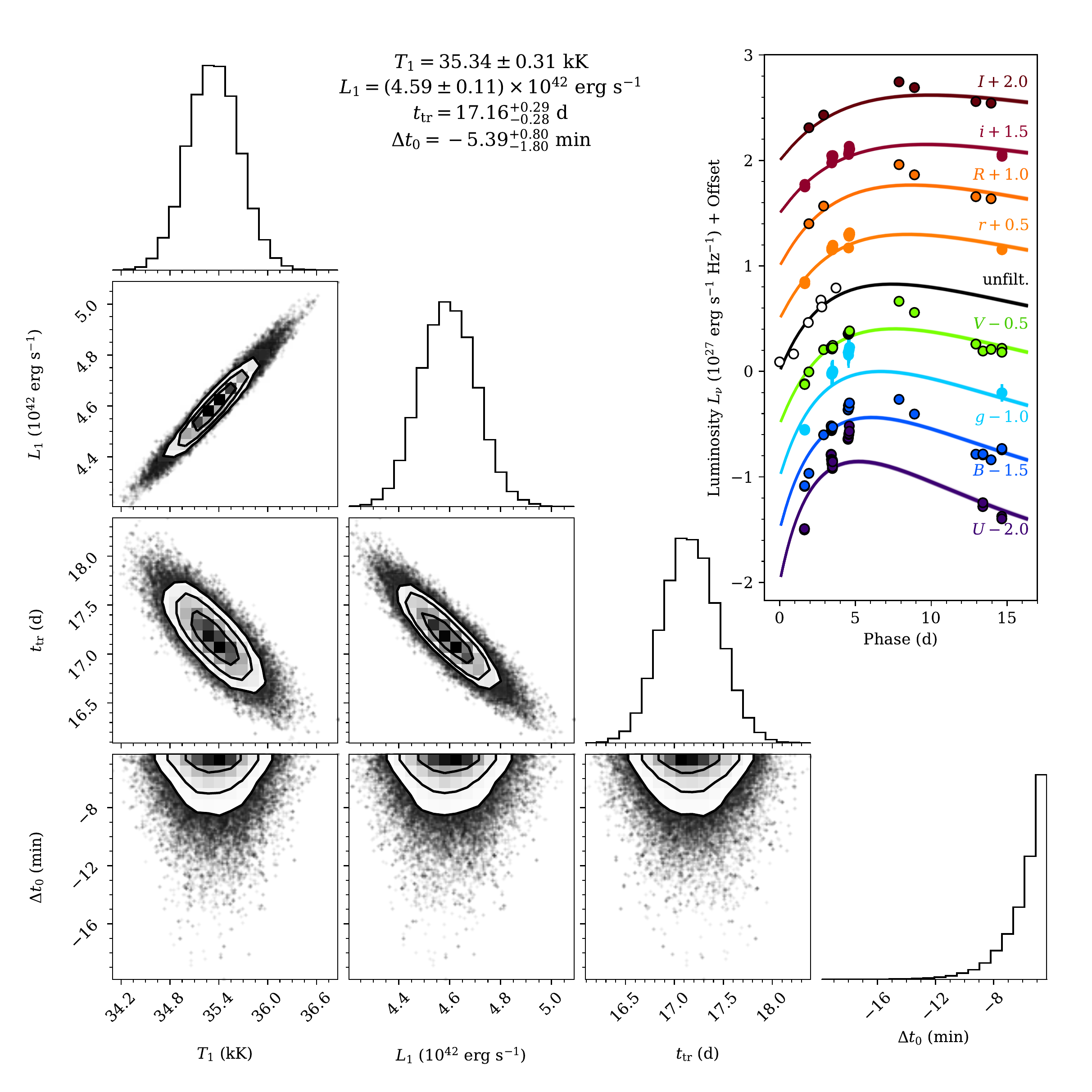}
\caption{\scriptsize Posterior probability distributions of and correlations between the temperature 1~day after explosion ($T_1$), the total luminosity $\approx$1~day after explosion ($L_1$), the time at which the envelope becomes transparent ($t_\mathrm{tr}$), and the time between explosion and discovery ($\Delta t_0$). The $1\sigma$ credible intervals centered around the median are given at the top. The top-right panel shows 100 fits randomly drawn from our MCMC routine compared to the data. (The fits appear to be single lines because the spread in the parameters is small.) We do not consider the shock cooling models to be a good fit to the light-curve peak, leaving circumstellar interaction as a possible additional power source.\label{fig:lcfits}}
\end{figure*}

We fit our multiband light curve up to MJD 57485.0 (16.3 days after estimated explosion) to the \citeauthor{Sapir2017} model, with $n=1.5$ for an RSG, using a Markov chain Monte Carlo (MCMC) routine and flat priors for all parameters. This results in posterior probability distributions for the temperature 1~day after explosion ($T_1$), the total luminosity $\approx$1~day after explosion ($L_1$), the time at which the envelope becomes transparent ($t_\mathrm{tr}$), and the time of explosion ($t_0$). For each set of these parameters, the model gives the blackbody temperature and total luminosity as a function of time, which we then convert to observed fluxes for each photometry point, simultaneously fitting all bands. Figure~\ref{fig:lcfits} shows the light-curve fits, posterior probability distributions, and $1\sigma$ credible intervals centered on the medians.

\citeauthor{Sapir2017}'s models are valid starting when the supernova shock traverses the progenitor radius ($t > \frac{R}{5v}$) and ending when the ejecta cool below 8000~K. For our best-fit parameters, the latter occurs 27~days after explosion, well after the range of points we fit, but the former occurs 1--2~days after explosion. The most notable feature of our fits is that the explosion time is pushed as late as possible (5~minutes before discovery). If we believe this value, we should exclude the first two unfiltered points from our fit. However, this is in conflict with our earlier estimate of the explosion date, and as we discuss below, we do not consider these models to be a good fit.

\begin{figure*}
\plotone{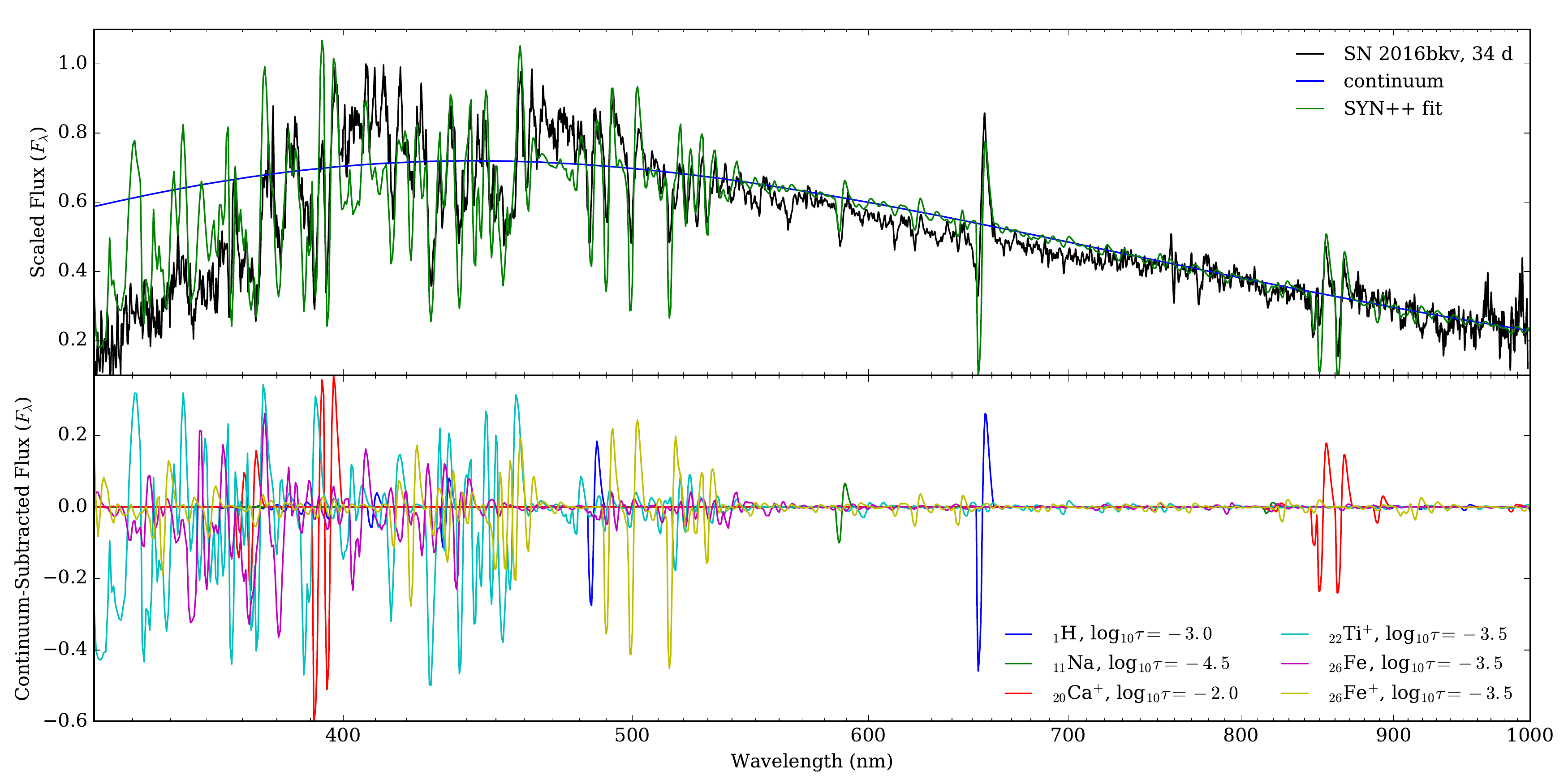}
\caption{\scriptsize SYN++ model for the spectrum of SN~2016bkv 34.7 days after maximum light. The top panel shows the observed spectrum, the assumed continuum (a 7500~K blackbody warped by a quadratic polynomial; see Section~\ref{sec:synapps}), and the total synthetic spectrum (also warped). The bottom panel shows the contributions to the synthetic spectrum of each of the six ions we consider. (The SYN++ input file used to create this figure is available.)\label{fig:synapps}}
\end{figure*}

Although the fit appears to converge on an optimal set of parameters (see also Figure~\ref{fig:bb}), the models are not able to reproduce the sharpness of the peak around 8~days after explosion. This may be evidence that ejecta-CSM interaction contributes significant luminosity during the week around maximum light, which we might expect given our observation of the presence of CSM through flash spectroscopy. Indeed, \cite{Morozova2017,Morozova2018} have shown that numerical light-curve models that include CSM reproduce these bumps much more faithfully than those with no CSM. In fact, there may be a correlation between the strength of the flash spectroscopy features and the luminosity in the initial light-curve peak, but more events with both types of measurements are needed to investigate such a correlation further.

\subsection{Spectral Modeling}\label{sec:synapps}
Because of their low photospheric velocities, LL SNe~IIP provide a good opportunity to identify which elements contribute to their photospheric spectra. To aid in this identification, we use SYN++ \citep{Thomas2011} to produce a synthetic spectrum that resembles our observed spectrum from 34.7 days after explosion. We chose this spectrum because it has a high signal-to-noise ratio, and many strong P~Cygni lines are visible. Figure~\ref{fig:synapps} compares the synthetic and observed spectra. The synthesis parameters are listed in the SYN++ input file, which is available in the online journal.

At this phase, the blue half of the optical spectrum is dominated by iron and titanium lines. These two elements plus calcium, sodium, and hydrogen can account for nearly all of the observed features. The most notable feature of our result is that the range of input velocities necessary to reproduce the observed spectrum is only 100--2000~km~s$^{-1}$. Likewise, the photospheric velocity is 1000~km~s$^{-1}$. Since this is near the resolution of our spectrograph, the true ejecta velocity may be lower. The photospheric temperature of our model is 7500~K, close to the blackbody temperature we calculate for this phase in Section~\ref{sec:bolometric}, and the ion temperatures of $10^4$~K are reasonable for the interior of the ejecta (although the synthetic spectrum is not very sensitive to these parameters).

Although the line positions match the observed spectrum quite well, a quadratic warping function is necessary to suppress the blue end of the model to match the observations. This function has no physical basis, but deviations from a perfect blackbody continuum with no electron scattering are conceivable. Another possibility is that our spectrograph has lost some of the blue light from the supernova in a way we have not sufficiently accounted for.

\subsection{Nebular Spectra}

Nebular spectra, taken after the ejecta are mostly transparent to optical light, provide several clues about the progenitor structure. We obtained four late-time spectra of SN~2016bkv, from 257 to 607~days after explosion. Because the SN was faint at these phases, the spectra are significantly contaminated by light from the host \ion{H}{2} region. No spectrum of this \ion{H}{2} region exists, so for each SN spectrum, we simply modeled the continuum using a Gaussian process with a kernel width of 100~nm and subtracted it from the observed spectrum. (Note, however, that this will not remove the contamination of the hydrogen lines.) We then scaled the result by a constant factor to match an extrapolation of the host-subtracted $r$- and $i$-band photometry.

\begin{figure*}
\centering
\includegraphics[width=\textwidth]{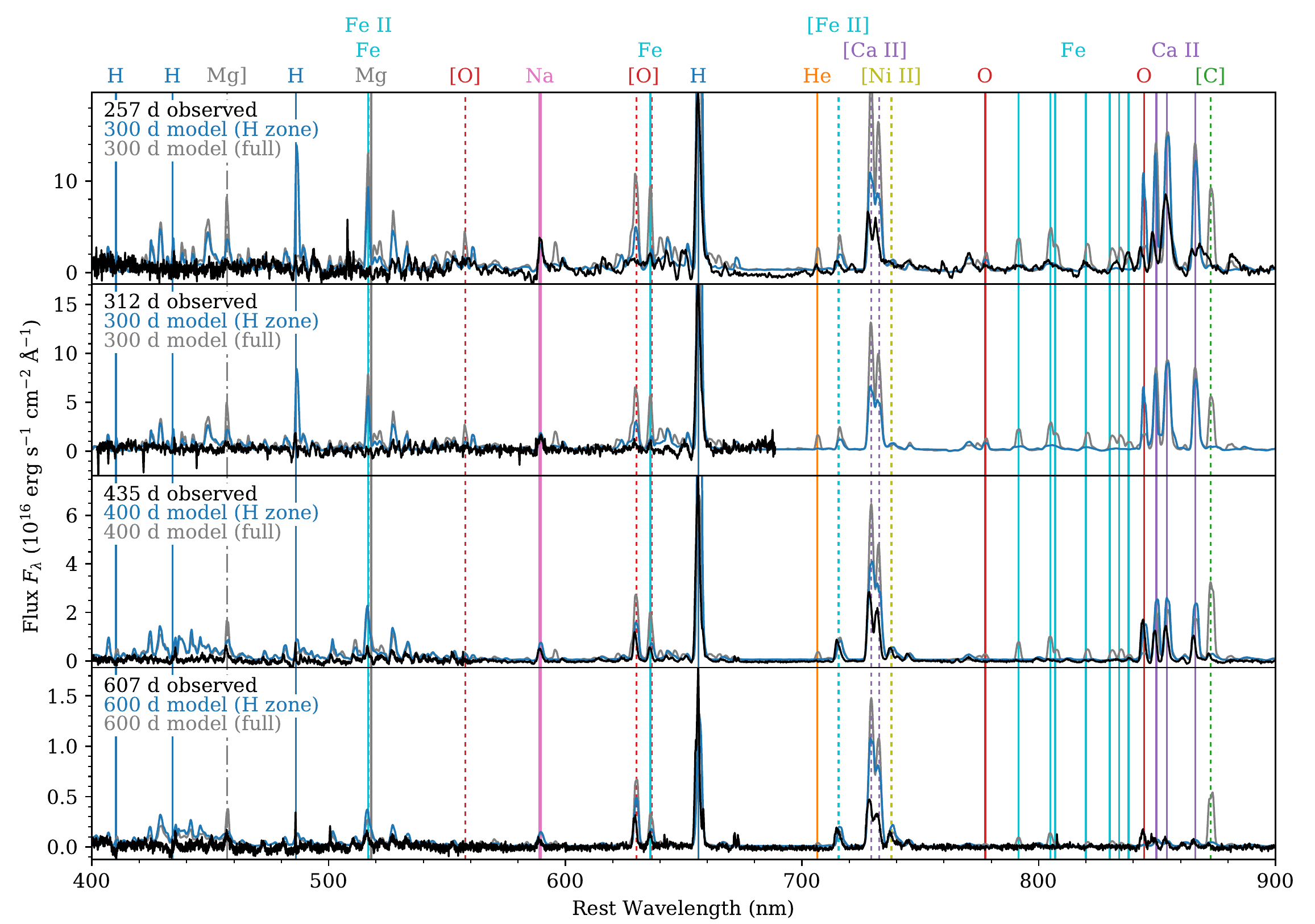}
\caption{\scriptsize Late-time spectra of SN~2016bkv (black) compared to the $9\,M_\sun$ model of \cite{Jerkstrand2018}: the full model in gray and hydrogen-zone model in blue. The observed spectra are calibrated to extrapolated $r$- and $i$-band photometry, and the model spectra are scaled to match the nickel mass and phase of the observed spectra. The agreement is not perfect, but there is a notable similarity between the observed spectra and the hydrogen-zone model, especially around 400~days. This raises the possibility that SN~2016bkv may be an electron-capture supernova from a low-mass RSG progenitor, although more specific models are required before we can make a firm conclusion.\label{fig:nebular}}
\end{figure*}

In Figure~\ref{fig:nebular}, we compare these to the nebular spectral models of \cite{Jerkstrand2018} for a $9\,M_\sun$ RSG progenitor. These models were calculated using solar metallicity, which is not necessarily applicable for this \ion{H}{2} region, but the literature contains very few other models that are appropriate for LL SNe~II \citep[e.g.,][]{Lisakov2017,Lisakov2018}. The model spectra have been scaled to match the nickel luminosity of SN~2016bkv at the observed phase. Although \citeauthor{Jerkstrand2018} do not model an electron-capture supernova, they expect nebular spectra from such an explosion to resemble their ``pure hydrogen-zone'' model, in which the progenitor is composed entirely of material from the hydrogen envelope (see their Figure~2). Specifically, \ion{Mg}{1}], \ion{Fe}{1}, [\ion{O}{1}], \ion{He}{1}, [\ion{C}{1}], and \ion{O}{1} 777.4~nm would be weaker to nonexistent and \ion{O}{1} 844.6~nm would be stronger in the hydrogen-zone model. We plot this zone of the model in blue in Figure~\ref{fig:nebular}.

There are several discrepancies between the observed spectra and the full model, particularly the absence of observed iron lines, but the agreement between the observed spectra and the hydrogen-zone model is significantly better. As such, we consider the possibility that SN~2016bkv is an electron-capture supernova, although this conclusion is far from certain. An important caveat is that electron-capture supernovae are expected to yield very little radioactive nickel. On one hand, our spectra do not show any unambiguous lines from iron, the decay product of $^{56}$Ni, but on the other hand, we inferred an unusually large mass of $^{56}$Ni to power the late-phase light curve in Section~\ref{sec:bolometric}. Furthermore, \cite{Moriya2014} expect electron-capture supernovae to come from super-asymptotic giant branch stars, in which case they would explode inside denser and more massive wind than we infer here. More observations and further modeling of LL SNe~IIP in the nebular phase will be critical to understanding which, if any, supernovae are caused by electron capture.

\section{Conclusions}
Two lines of evidence point to strong but short-lived circumstellar interaction in SN~2016bkv. First, flash spectroscopy during the first five days reveals the presence of material around the progenitor star. A few days later, the light curve shows a strong peak that cannot be fit with the \cite{Sapir2017} shock cooling models alone, where the excess could represent luminosity from ejecta-CSM interaction. Even with this strong early peak at $M_V = -16$, SN~2016bkv is the lowest-luminosity supernova to show flash ionization lines, suggesting that late-stage mass loss is common even among LL SN~II progenitors.

SN~2016bkv is also exceptional in its short fall from plateau, indicating a large nickel production ($0.022\,M_\sun$) compared to other LL SNe~II. Certain emission lines in the nebular spectra of SN~2016bkv hint at the possibility of its being an electron-capture supernova, although we cannot yet definitively distinguish between this case and the traditional iron core-collapse model.

This and other work suggests that analyses of SNe~II should be careful to distinguish between the properties of the progenitor star itself and the properties of its circumstellar environment. Peak luminosity, for example, can be strongly affected by circumstellar interaction, whereas luminosity after settling on the plateau may be related to the intrinsic properties of the star. Future data sets like the one presented here, with early and long-term coverage of young supernovae, will allow us to constrain progenitor properties and mass-loss history separately, by comparing to numerical light-curve and spectral models.

\acknowledgements
We thank Nir Sapir for his advice on fitting the shock cooling models and Rollin Thomas for his help using SYN++. We are grateful to the staff at the Keck Observatory for their assistance. The W.~M.~Keck Observatory is operated as a scientific partnership among the California Institute of Technology, the University of California, and NASA; it was made possible by the generous financial support of the W.~M.~Keck Foundation. Research at Lick Observatory is partially supported by a generous gift from Google.

G.H., C.M., and D.A.H.\ are supported by the National Science Foundation under grant No.~1313484. Support for I.A.\ was provided by NASA through the Einstein Fellowship Program. X.W.\ is supported by the National Natural Science Foundation of China (NSFC grants 11325313 and 11633002). This work was also partially Supported by the Open Project Program of the Key Laboratory of Optical Astronomy, National Astronomical Observatories, Chinese Academy of Sciences. G.H.\ thanks the LSSTC Data Science Fellowship Program; his time as a Fellow has benefited this work.

\facilities{ADS, Beijing:0.8m, Beijing:2.16m (BFOSC, OMR), Keck:I (LRIS), LCOGT (FLOYDS, Sinistro), NED, Shane (Kast)}.
\defcitealias{AstropyCollaboration2018}{Astropy Collaboration 2018}
\software{Astropy \citepalias{AstropyCollaboration2018}, \texttt{corner.py} \citep{corner}, \texttt{emcee} \citep{Foreman-Mackey2013}, \texttt{george} \citep{Ambikasaran2015}, HOTPANTS \citep{HOTPANTS}, \texttt{lcogtsnpipe} \citep{Valenti2016}, Matplotlib \citep{matplotlib}, NumPy \citep{numpy}, PyRAF \citep{PyRAF}, PyZOGY \citep{PyZOGY}, SExtractor \citep{Bertin1996}, SYN++ \citep{Thomas2011}}.

\bibliography{zotero_abbrev,software}

\end{document}